\documentclass[fleqn,draf]{article}

\usepackage[utf8]{inputenc}
\usepackage[T1]{fontenc}

\usepackage[a4paper,margin=2cm]{geometry}

\usepackage{mathtools,amsthm}
\newtheorem{theorem}{Theorem}
\usepackage{mdframed}

\usepackage[dvipsnames]{xcolor} 
\usepackage{hyperref,cleveref}
\hypersetup{
    colorlinks,
    linkcolor=Blue,
    citecolor=Blue
}

\usepackage{graphicx}
\usepackage{subfig}

\def\half{\tfrac{1}{2}}

\usepackage[per-mode=symbol]{siunitx}

\usepackage{authblk}

\begin{document}

\title{Response of an Interferometer Mounted on an Elastic Square Plate to Gravitational Waves}
\author[1]{Thomas Spanner}
\author[2]{Thomas B.\ Mieling%
\thanks{ORCID:~\href{https://orcid.org/0000-0002-6905-0183}{\texttt{0000-0002-6905-0183}}}%
}
\author[1]{Stefan Palenta%
\thanks{ORCID:~\href{https://orcid.org/0000-0002-6541-9537}{\texttt{0000-0002-6541-9537}}}%
}
\affil[1]{University of Vienna, Faculty of Physics, Boltzmanngasse 5, 1090 Vienna, Austria}
\affil[2]{University of Vienna, Faculty of Physics, Vienna Doctoral School in Physics (VDSP), Vienna Center for Quantum Science and Technology (VCQ) and Research platform TURIS, Boltzmanngasse~5, 1090 Vienna, Austria}
\maketitle

\begin{abstract}
    Laser-interferometric gravitational wave detectors are commonly modeled as being at rest in transverse-traceless coordinates (and thus geodesic).
    In this paper, we analyze what happens if the interferometer is mounted on a material that can undergo elastic oscillations caused by the gravitational wave.
    We thus compute the response of a two-dimensional elastic material to linearized gravitational radiation and compute the resulting response of a laser interferometer, mounted on such a plate.
\end{abstract}

\tableofcontents

\section{Introduction}
This work builds upon the paper \cite{2023CQGra..40h5007H} by Hudelist \textit{et al}. Therein, the equations of motion describing an elastic body under the influence of gravitational waves (GWs) were derived by using a concrete matter model in the theory of relativistic elasticity. These were then used to solve the simple one-dimensional problem of a rod in a GW background. This paper aims to generalize this to a two-dimensional thin plate and then goes on to calculate the signal a Michelson interferometer placed on the plate would measure.
In other words, it derives the response of elastic interferometers to a GW.
This refines previous models of laser-interferometric GW detection in media, such as the ones in Refs.~\cite{2021CQGra..38o5006M,2021CQGra..38u5004M}.
The model considered here does not include any damping behavior and we only search for the steady-state solution, when a continuous GW hits the plate.

First, we discuss a set of normal modes, i.e.\ solutions for the in-plane vibrations of the plate without a gravitational wave. Next, we express the solution as a Fourier series, plug it into the equations, and solve for the Fourier coefficients. But for non-periodic functions, taking the derivative of the Fourier series does not result in the Fourier series of the derivative. To fix this, we develop a modified spectral approach which is then used to find a solution. The resulting series is truncated, and the linear system of equations is then solved numerically.
This is then used to compute the signal in a Michelson interferometer, whose constituents (the laser, beam splitter, and mirrors) are attached to the elastic plate. Finally, the results are compared to those obtained for an interferometer whose constituents are at rest in transverse-traceless coordinates.

We use the metric signature $(-,+,+,+)$, Greek letters denote spacetime indices and Latin letters denote spacial indices.
Furthermore, we use units with the speed of light in vacuo set to unity: $c=1$.

\section{Elastic Plate in the Presence of Gravitational Waves}

Far from the source, GWs are weak and therefore can be described using linearized gravity
\begin{equation}
    g_{\mu \nu} = \eta_{\mu \nu} + \epsilon h_{\mu \nu} \,,
\end{equation}
with $\epsilon \ll 1$. In the transverse-traceless (TT) gauge, a monochromatic plane wave shall be expressed via
\begin{equation}
    h_{ij} = \Re 
    \begin{pmatrix}
        A_+ & A_\times & 0 \\
        A_\times & -A_+ & 0 \\
        0 & 0 & 0
    \end{pmatrix} e^{i \omega (t-z)} \,.
\end{equation}
We wish to determine the deformation of homogeneous and isotropic materials due to such a gravitational wave. Within the framework of linearized relativistic elasticity (cf.\ e.g.\ \cite{2023CQGra..40h5007H}), the deformation is encoded in the displacement field $u^i$, which gives rise to the strain
\begin{equation}
    e_{ij}
     = \half(\partial_i u_j + \partial_j u_i + h_{ij})\,,
\end{equation}
and the Cauchy stress tensor
\begin{equation}
    \sigma^{ij}
        = \lambda \delta^{ij} e^k_k
        + 2 \mu e^{ij}
        + O(\epsilon^2)\,,
\end{equation}
where $\lambda$ and $\mu$ are the Lamé parameters of the material.
Furthermore, the stress-strain relation for an isotropic and homogeneous body (cf.\ e.g.\ \cite{2023CQGra..40h5007H}) reads
\begin{equation} \label{stress-strain}
    \sigma^{ij} = \lambda \delta^{ij} e^k_k+ 2 \mu e^{ij} + O(\epsilon^2) \,.
\end{equation}
Here, we restrict the discussion to the case where the material under consideration is a thin plate, lying in the $z = 0$ plane.
The absence of tensions normal to the plate is then modeled by setting $\sigma^{iz}=0$. Therefore, using \eqref{stress-strain} the strain tensor $e$ can be expressed in terms of its $x$- and $y$-components as
\begin{equation}
    e = \begin{pmatrix}
        e^{11} & e^{12} & 0 \\
        e^{12} & e^{22} & 0 \\
        0 & 0 & -\frac{\lambda}{\lambda + 2 \mu}(e^{11}+e^{22})
    \end{pmatrix}\,.
\end{equation}
The equations of motion of linearized elasticity then yield
\begin{equation}\label{eq:EoMFullGW}
    \rho \partial_t^2 u^j
        = \partial_i \sigma^{ij}
        =  \mu \frac{3 \lambda + 2\mu}{\lambda+2\mu} \partial^j \partial_i u^i + \mu \Delta u^j
       \,.
\end{equation}
We seek the steady-state solution oscillating with the GW frequency $\omega$. Therefore, we make the ansatz $u^i = \cos(\omega t) \varphi^i(x,y)$. Plugging this into \cref{eq:EoMFullGW} yields an equation only for $\varphi^i(x,y)$:
\begin{equation}
\label{eq:phi bulk equations}
    \omega^2 \varphi^i
    + c_2^2 \Delta \varphi^i
    + c_3^2 \partial^i \partial_k \varphi^k
    = 0 \,,
\end{equation}
where we have introduced the wave speeds $c_1^2 = \frac{4\mu}{\rho} \frac{\lambda + \mu}{\lambda+2\mu}$, $c_2^2 = \frac{\mu}{\rho}$ and $c_3^2 = c_1^2 -
c_2^2$ (cf.\ \cref{sec:FourierAnsatz} for further discussions). Note that this is consistent with the more convenient representation $c_1^2 = \frac{E}{\rho(1-\nu^2)}$ of the longitudinal wave speed $c_1$ using the Young modulus $E$ and the Poisson ratio $\nu$ \cite{Landau}.

For a quadratic plate of length $L$ in the $z = 0$ plane, these equations must be solved on the domain $P = \{(x,y)\rvert x,y \in [-L/2, +L/2]\}$.

The bulk equations of motion must be supplemented by boundary conditions on the boundary $\partial P$.
In the absence of external forces, one has $\sigma^{ij} n_j |_{\partial P} = 0$, where $n$ is the unit outward-pointing normal to $\partial P$ (within a time-slice of constant $t$) \cite{2023CQGra..40h5007H}.
Splitting the boundary $\partial P$ as $\partial P = \partial P_x \cup \partial P_y$, where $\partial P_x =\{(x, y)|x=\pm L/2, y \in [-L/2, L/2]\}$, and $\partial P_y = \{(x, y)|y=\pm L/2, x \in [-L/2, L/2]\}$, one has
\begin{subequations}\label{eq:BoundaryConditionsPhi}
\begin{align}
    \left[
        c_1^2 \partial_x  \varphi^x
        + (c_1^2-2 c_2^2) \partial_y \varphi^y
    \right]\big|_{\partial P_x}
        &= -c_2^2 A_+ \,,
    \\
    \left[
        c_1^2  \partial_y  \varphi^y
        + (c_1^2-2 c_2^2) \partial_x \varphi^x
    \right]\big|_{\partial P_y}
        &= + c_2^2 A_+\,,
    \\
    \left[
        \partial_x  \varphi^y
        + \partial_y  \varphi^x
    \right]\big|_{\partial P}
      & = - A_\times\,.
\end{align}
\end{subequations}

\section{Delta Corrected Spectral Method}
\label{chap:Delta}

For clarity, within this section, we make use of the rescaling $x^i \to x^i/L$ to dimensionless coordinates and the corresponding dimensionless wave speeds $\Bar{c_i} = \tfrac{c_i}{\omega L}$. Then the differential equation \eqref{eq:phi bulk equations} turns into
\begin{equation} \label{eq:ReducedSpatialEq}
    \varphi^i + \Bar{c}_2^2 \Delta \varphi^i + \Bar{c}_3^2 \partial^i \partial_k \varphi^k = 0 \,.
\end{equation}
and the boundary conditions \eqref{eq:BoundaryConditionsPhi} read
\begin{equation} \begin{aligned} \label{eq:ReducedBoundaryConditions}
    \bar{c}_1^2 \partial_x  \varphi^x\rvert_{\partial P_x} + (\bar{c}_1^2-2 \bar{c}_2^2) \partial_y  \varphi^y\rvert_{\partial P_x}  & = -\bar{c}_2^2 L A_+ \,,\\
    \bar{c}_1^2  \partial_y  \varphi^y\rvert_{\partial P_y} + (\bar{c}_1^2-2 \bar{c}_2^2) \partial_x  \varphi^x\rvert_{\partial P_y}  & = + \bar{c}_2^2 L A_+ \,,\\
    \partial_x  \varphi^y\rvert_{\partial P} + \partial_y  \varphi^x\rvert_{\partial P}  & = - L A_\times \,.
\end{aligned} \end{equation}
We wish to implement a spectral method to (approximately) solve the boundary value problem (BVP) consisting of \cref{eq:ReducedSpatialEq,eq:ReducedBoundaryConditions}. However, since the solution $\varphi$ is in general not periodic, the derivatives taken from its Fourier series representation must be adapted to solve the differential equation \eqref{eq:ReducedSpatialEq}. In the first step, this procedure shall be introduced in a one-dimensional BVP.

\subsection{Illustration of the Problem in 1D}
\label{sec:1DProblem}
\begin{figure}[ht!]
    \centering
    \includegraphics[width=0.7\columnwidth]{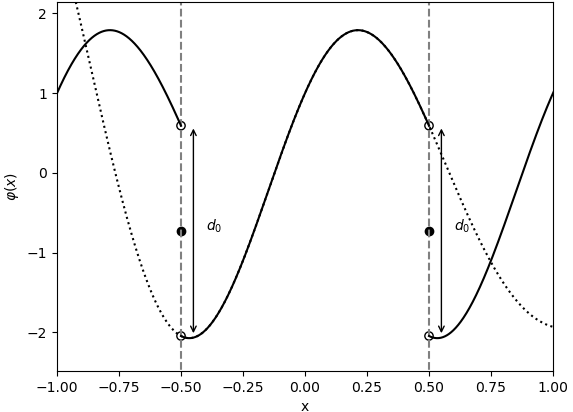}
    \caption{Periodic continuation (black) and original function (dotted)}
    \label{fig:PeriodicContf}
\end{figure}

The smooth solution $\varphi(x)$ of a one-dimensional BVP has the Fourier expansion
\begin{equation}
    F[\varphi](x)
        = \sum_{n=0}^{\infty} a_n \cos(2\pi n x) + \sum_{n=1}^{\infty} b_n \sin(2\pi n x)\,.
\end{equation}
But since that solution is not necessarily periodic on $[-\frac{1}{2}, \frac{1}{2}]$, the Fourier series actually represents the periodic continuation of $\varphi(x)$ (see \cref{fig:PeriodicContf}), which in general has a jump at the boundary. At such points, the Fourier series $F[\varphi]$ takes the average of both one-sided limits of the original function. The size of the jump shall be denoted by $d_0=\varphi(+\frac{1}{2})-\varphi(-\frac{1}{2})$. Then, the correct function values can be recovered by the following relation:
\begin{equation}
    \varphi(x) = \begin{cases}
    F[\varphi](x) & x \in (-\frac{1}{2},\frac{1}{2})\,, \\
    F[\varphi](\frac{1}{2}) \pm \half d_0 & x=\pm \frac{1}{2}\,.
    \end{cases}
\end{equation}
Now the derivative of the periodic continuation $\partial_x F[\varphi](x)$ differs from the periodic continuation of the derivative $F[\varphi']$ by a jump of height $-d_0$ at the boundary and all its periodic recurrences. This can be expressed via
\begin{equation}
    F[\varphi'] = \partial_x F[\varphi] + d_0 F[\delta(x-\half)]\,.
\end{equation}
The first derivative is again not necessarily periodic, and therefore has a jump $d_1=\varphi'(+\half)-\varphi'(-\half)$ at the boundary. So the above procedure can be iterated yielding
\begin{equation}
    F[\varphi''] = \partial_x F[\varphi'] + d_1 F[\delta(x-\half)]= \partial_x^2 F[\varphi] + d_0 F[\delta '(x-\half)] + d_1 F[\delta(x-\half)]\,.
\end{equation}
plugging this into the one-dimensional version of \eqref{eq:ReducedSpatialEq} and solving for the Fourier coefficients $\{a_n,b_n\}$ correctly reproduces the solution for a one-dimensional rod under the influence of a GW as found in Ref.~\cite{2023CQGra..40h5007H}.

\subsection{Fourier Series with Dirac Deltas in Two Dimensions}
\label{2DFourierDeltaSec}

In two dimensions we write the Fourier series of an arbitrary function $f(x,y)$ on the square $[-\half,+\half] \times [-\half,+\half]$ in terms of complex exponentials
\begin{align}
    F[f](x,y)
        &= \sum_{n,m=-\infty}^{\infty} C_{nm} e^{i \vec{k} \cdot \vec{x}}\,,
    &
    \vec{k}
        &= 2 \pi \begin{pmatrix} n \\ m \end{pmatrix}\,.
\end{align}
\begin{figure}[ht!]
    \centering
    \includegraphics[scale=0.7]{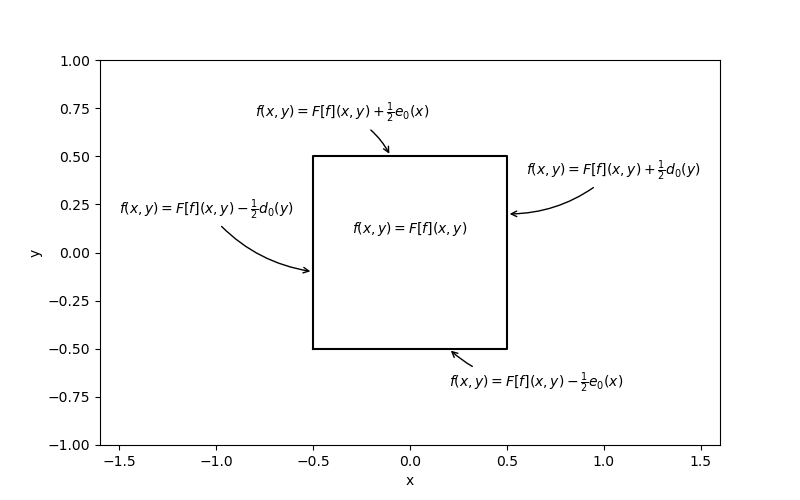}
    \caption{Relation between function at its Fourier series}
    \label{Fig2DFourierValues}
\end{figure}
Once again, the function is in general not periodic and so the Fourier series only agrees with the function inside the square (see \cref{Fig2DFourierValues}). Instead of a constant describing the jump size, there are now two jump functions, one along $\partial P_x$ (denoted by $d_0$) and one along $\partial P_y$ (denoted by $e_0$).
\begin{align}
    d_0(y) &= f( \half,y ) -f(-\half,y)\,,
    &
    e_0(x) &= f(x,\half)-f(x,-\half)\,.
\end{align}
To get the correct Fourier series for the partial derivatives of $f$ we need, similar to the one-dimensional case, to add the periodic continuation the Dirac $\delta$-function to the partial derivative of the Fourier series. For $\partial_x f$ this means:
\begin{equation}
    F[\partial_x f(x,y)] = \partial_x F[f(x,y)] + F[ d_0(y) \delta (x-\half ) ]\,.
\end{equation}
But now the jump is represented by a function, instead of a constant. To get back the correct values of the derivatives at the boundaries we again need jump functions:
\begin{align}
    d_x(y) &= \partial_x f ( \half,y ) - \partial_x f(-\half,y) \,, \\
    e_x(x) &= \partial_x f(x,\half)- \partial_x f(x,-\half)\,.
\end{align}
The subscript denotes the derivative to which the jump function belongs, e.g.\ $d_x$ for the jump in $\partial_x f$. The relation between the function values and the Fourier series is then given by
\begin{equation}
    \partial_x f(x,y)
    = \begin{cases}
        F[\partial_x f(x,y)]
            & - \frac{1}{2} < x,y < \frac{1}{2} \,, \\
        F[\partial_x f(x,y)] \pm \frac{1}{2} d_x(y)
            & x = \pm \frac{1}{2}, - \frac{1}{2} < y < \frac{1}{2} \,, \\
        F[\partial_x f(x,y)] \pm \frac{1}{2} e_x(x)
            & y = \pm \frac{1}{2}, - \frac{1}{2} < x < \frac{1}{2} \,.
    \end{cases}
\end{equation}
To get the second derivatives we treat the first derivative as the function $f$ and use what we already know:
\begin{equation}\begin{aligned}
    F[\partial_x^2 f(x,y)] & = \partial_x F[\partial_x f(x,y)] + F[ d_x(y) \delta ( x-\half ) ] \\
    & = \partial_x \left( \partial_x F[f(x,y)] + F[ d_0(y) \delta (x-\half ) ] \right) + F[ d_x(y) \delta ( x-\half ) ] \\
    & =\partial_x^2 F[f(x,y)] + F[ d_0(y) \delta' (x-\half ) ]  + F[ d_x(y) \delta ( x-\half ) ] \,,
\end{aligned}\end{equation}
The second $y$-derivative looks very similar to the expression in the 1D case, but the mixed derivative is more interesting, because the second derivative now also acts on the jump functions.
\begin{equation}
    \begin{split}
        F[\partial_x \partial_y f] 
        &= \partial_x \partial_y F[f]
        + \partial_x F[ e_0(x)] F[ \delta (y-\half ) ] \\
        &+ \partial_y F[ d_0(y) ] F[\delta ( x-\half ) ] 
        + J F[\delta (x-\half ) \delta (y-\half )] \,.
    \end{split}
\end{equation}
Here, $J$ denotes the jump in the jump functions:
\begin{equation*}
    J := d_0(\half)-d_0(-\half) = f(\half,\half) - f(-\half, \half) - f(\half,-\half) + f(-\half,-\half) = e_0(\half)-e_0(-\half) \,.
\end{equation*}
The desired solution to \cref{eq:ReducedSpatialEq} with the boundary conditions given in \cref{eq:ReducedBoundaryConditions} has an $x$- and $y$-component and is called $\varphi$ instead of $f$. Therefore, all Fourier coefficients and jump functions also get an index.
It is useful to first look at the boundary conditions expressed in terms of the Fourier series. For instance, the expression for $\sigma_{xy}$ at $x=\pm \half$ reads
\begin{equation}
    F[\partial_x \varphi^y(\pm \half,y)] \pm \half d_x^y(y) + F[\partial_y \varphi^x(\pm \half,y)] \pm \half d_y^x(y) + A_\times = 0 \,.
\end{equation}
The Fourier series at $x=\half$ has the same value as the one at $x=-\half$ so when both cases are subtracted from one another what remains is the relation 
\begin{equation}
    d^y_x(y) + d_y^x(y) = 0 \,.
\end{equation}
Similar relations can be found when considering $\sigma_{xx}$ on $\partial P_x$, $\sigma_{yy}$ on $\partial P_y$ and $\sigma_{xy}$ on $\partial P_y$ respectively:
\begin{align}
    \bar{c}_1^2 d^x_x(y) + (\bar{c}_1^2- 2 \bar{c}_2^2) d_y^y(y) &= 0 \,, \\
   (\bar{c}_1^2- 2 \bar{c}_2^2) e^x_x(x) + \bar{c}_1^2 e_y^y(x) &= 0 \,, \\
    e^y_x(x) + e_y^x(x) &= 0 \,.
\end{align}
Using these, all equations can be expressed in terms of $d^i_0$ and $e^i_0$ only, e.g.
\begin{equation}
    d_x^x = -\tfrac{\bar{c}_1^2- 2 \bar{c}_2^2}{\bar{c}_1^2} d_y^y = -\tfrac{\bar{c}_1^2- 2 \bar{c}_2^2}{\bar{c}_1^2} \partial_y d_0^y \,.
\end{equation}
For the second $x$- and $y$-derivatives this means
\begin{align*}
    F[\partial_x^2 \varphi^x] &= \partial_x^2 F[\varphi^x] + F[ d_0^x(y) \delta' (x-\half ) ]  - \tfrac{\bar{c}_1^2- 2 \bar{c}_2^2}{\bar{c}_1^2} F[ \delta ( x-\half ) ] \left( \partial_y F[ d_0^y(y)]  + J F[\delta (y-\half )] \right) \,, \\
    F[\partial_y^2 \varphi^x] &= \partial_y^2 F[\varphi^x] + F[ e^x_0(y) \delta' (x-\half ) ]  - \partial_x F[ e_0^y(x)] F[ \delta \left( x-\half \right) ] - J F[\delta (x-\half ) \delta (y-\half )] \,.
\end{align*}
Each of the spatial derivatives in \cref{eq:ReducedSpatialEq} includes one term containing the factor $F[\delta (x-\half ) \delta (y-\half )]$. Looking only at their prefactors, one finds that these cancel each other and the Fourier series form of the equation becomes
\begin{equation*}
\begin{aligned}
    F[\varphi^x] &+ \Bar{c}_1^2 \left( \partial_x^2 F[\varphi^x] + F[ d_0^x(y) \delta' (x-\half ) ]    \right) + (\bar{c}_1^2-2 \bar{c}_2^2)  \partial_x F[ e^y_0(x)] F[ \delta (y-\half ) ] \,,  \\
    &+ \Bar{c}_3^2  \partial_x \partial_y F[\varphi^y]  + \Bar{c}_2^2 \left( \partial_y^2 F[\varphi^x] + F[ e^x_0(x) \delta' (y-\half ) ] + \partial_y F[ d_0^y(y) ] F[\delta ( x-\half ) ]  \right)  = 0 \,.
\end{aligned}
\end{equation*}
One obtains a similar expression for the $y$ component of the equation.
These can now be turned into equations for the Fourier coefficients $C^i_{nm}$ of the function and the jumps $c_m(d^i_0)$ and $c_n(e^i_0)$. The same can be done with the boundary conditions which give four more expressions. 

So we now have six relations (for each $n$ and $m$) which contain the same information as the original BVP, but are expressed in Fourier coefficients. These are summarized in the following box.
\begin{theorem}
    In terms of the Fourier coefficients $C^j_{nm}$ for $\varphi^j$ and the Fourier coefficients $c_n(e_0^j)$ and $c_m(d_0^j)$ of the jump functions, the system consisting of \cref{eq:phi bulk equations,eq:BoundaryConditionsPhi} takes the form
    \begin{subequations}
    \label{eq:Fourier system}
    \begin{align}
        \label{eq:Fourier system a}
        \begin{split}
            &C_{nm}^x (1 - 4 \pi^2 (\Bar{c}_1^2 n^2 + \Bar{c}_2^2 m^2))
            - C_{nm}^y 4 \pi^2 n m \Bar{c}_3^2 \\
            &+ 2 i \pi [
                \Bar{c}_1^2 c_m(d_0^x) n (-1)^n
                + \Bar{c}_2^2 c_n(e_0^x)  m (-1)^m\\
                &\qquad\quad
                + (\Bar{c}_1^2 - 2 \Bar{c}_2^2)  n c_n(e_0^y) (-1)^m 
                + \Bar{c}_2^2  m c_m(d_0^y) (-1)^n 
            ] = 0\,,
        \end{split}
        \\
        \label{eq:Fourier system b}
        \begin{split}
            &C_{nm}^y (1 - 4 \pi^2 (\Bar{c}_2^2 n^2 + \Bar{c}_1^2 m^2))
            - C_{nm}^x 4 \pi^2 n m \Bar{c}_3^2 \\
            &+ 2 i \pi [
                \Bar{c}_2^2 c_m(d_0^y) n (-1)^n
                + \Bar{c}_1^2 c_n(e_0^y)  m (-1)^m
                \\&\qquad\quad
                + \Bar{c}_2^2  n c_n(e_0^x) (-1)^m
                + (\Bar{c}_1^2 - 2 \Bar{c}_2^2) m c_m(d_0^x) (-1)^n
            ] = 0\,,
        \end{split}
        \\
        \label{eq:Fourier system c}
        \begin{split}
            &\sum_{n=-\infty}^\infty (-1)^n
            [
                c_1^2 (2 \pi i n C_{nm}^x
                + c_m(d_0^x) (-1)^n)
                \\&\qquad\qquad\qquad
                + (c_1^2-2 c_2^2) (2 \pi i m C_{nm}^y + c_n(e_0^y) (-1)^m)
            ]
            = - c_2^2 L A_+ \delta_{m0}\,,
        \end{split}
        \\
        \label{eq:Fourier system d}
        \begin{split}
            &\sum_{m=-\infty}^\infty (-1)^m
            [
                c_1^2 (2 \pi i m C_{nm}^y + c_n(e_0^y) (-1)^m)
                \\&\qquad\qquad\qquad
                + (c_1^2-2 c_2^2) (2 \pi i n C_{nm}^x + c_m(d_0^x) (-1)^n)
            ]
            = + c_2^2 L A_+ \delta_{m0}\,,
        \end{split}
        \\
        \label{eq:Fourier system e}
        \begin{split}
            \sum_{n=-\infty}^\infty (-1)^n
            [
                2 \pi i m C_{nm}^x
                + c_n(e_0^x) (-1)^m
                + 2 \pi i n C_{nm}^y 
                + c_m(d_0^y) (-1)^n
            ]
            = - L A_\times \delta_{m0}\,,
        \end{split}
        \\
        \label{eq:Fourier system f}
        \begin{split}
        \sum_{m=-\infty}^\infty (-1)^m
            [
                2 \pi i m C_{nm}^x
                + c_n(e_0^x) (-1)^m
                + 2 \pi i n C_{nm}^y + c_m(d_0^y) (-1)^n]
            = - L A_\times \delta_{m0}\,.
        \end{split}
    \end{align}
    \end{subequations}
\end{theorem}

The recipe to solve these expressions is as follows: First, solve the first two relations for $C_{nm}^i$ in terms of the Fourier coefficients of the jump functions. Then plug these into the relations from the boundary conditions. Truncate the infinite series at some large value $M$ and solve the resulting linear system numerically.
The fact that there are as many equations as there are unknowns suggests the existence of a unique solution. This is confirmed by the numerical analysis presented below.
Then, the relations from the first step can be used to calculate the $C_{nm}^i$ which in turn can be used to approximate the solution via the truncated Fourier series.

\section{Physical Distances}\label{Phys_Dist}

\subsection{Coordinate Transformation to Local Lorentz Coordinates}

We want to mount an interferometer on the plate to measure gravitational waves. To calculate the signal it would measure we need to calculate the physical path lengths of the two interferometer arms. This is done using local Lorentz coordinates ($\xi, \eta$) \cite{Rakhmanov}, given by the transformation
\begin{align} \begin{split}
      \xi
        &= x + \half \epsilon A_+ x \cos(\omega t)+ \half \epsilon A_\times y \cos(\omega t) \,, \\
      \eta
        &= y - \half \epsilon A_+ y \cos(\omega t) + \half \epsilon A_\times x \cos(\omega t) \,,
\end{split} \end{align}
where $x$ and $y$ denote the TT coordinates as above.
In local Lorentz coordinates (LL coordinates) $\xi$ and $\eta$, physical distances are given by the difference in coordinates.

The TT-coordinates are related to the deformation by
\begin{equation}
    x^i = X^i + \epsilon u^i \,,
\end{equation}
where the upper case coordinates are the body coordinates $X^i$ (cf.\ \cite{2023CQGra..40h5007H}). Putting this all together and ignoring $\epsilon^2$ terms, we get
\begin{align} \begin{split} \label{PhysicalCoordTrafo}
    \xi
        &= X (1 + \half \epsilon A_+ \cos(\omega t)) +\half \epsilon A_\times  Y \cos(\omega t) + \epsilon u^x \,,  \\
    \eta
        &= Y (1 - \half \epsilon A_\times \cos(\omega t)) + \half \epsilon A_\times X \cos(\omega t) + \epsilon u^y \,.
\end{split}\end{align}

\subsection{Physical Displacement Field}
We can use the coordinate transformation \eqref{PhysicalCoordTrafo} to translate the displacement fields $(u^x, u^y)$, or their time-independent part $(\varphi^x, \varphi^y)$, into the physical displacement fields $(u^\xi, u^\eta)$ or $(\varphi^\xi, \varphi^\eta)$:
\begin{align}
    \label{eq:PhysicalDisplacementFields}
    \varphi^\xi
        &= \varphi^x + \half A_+ X + \half A_\times Y \,,
    &
    \varphi^\eta
        &= \varphi^y - \half A_+ Y  + \half A_\times X \,. 
\end{align}
For illustration, the plate’s material parameters are taken to be $c_1 = \SI{1950}{\metre\per\second}$ and $c_2= \SI{540}{\metre\per\second}$, which corresponds to polyethylene.

\begin{figure} 
    \centering
    \includegraphics[scale=0.7]{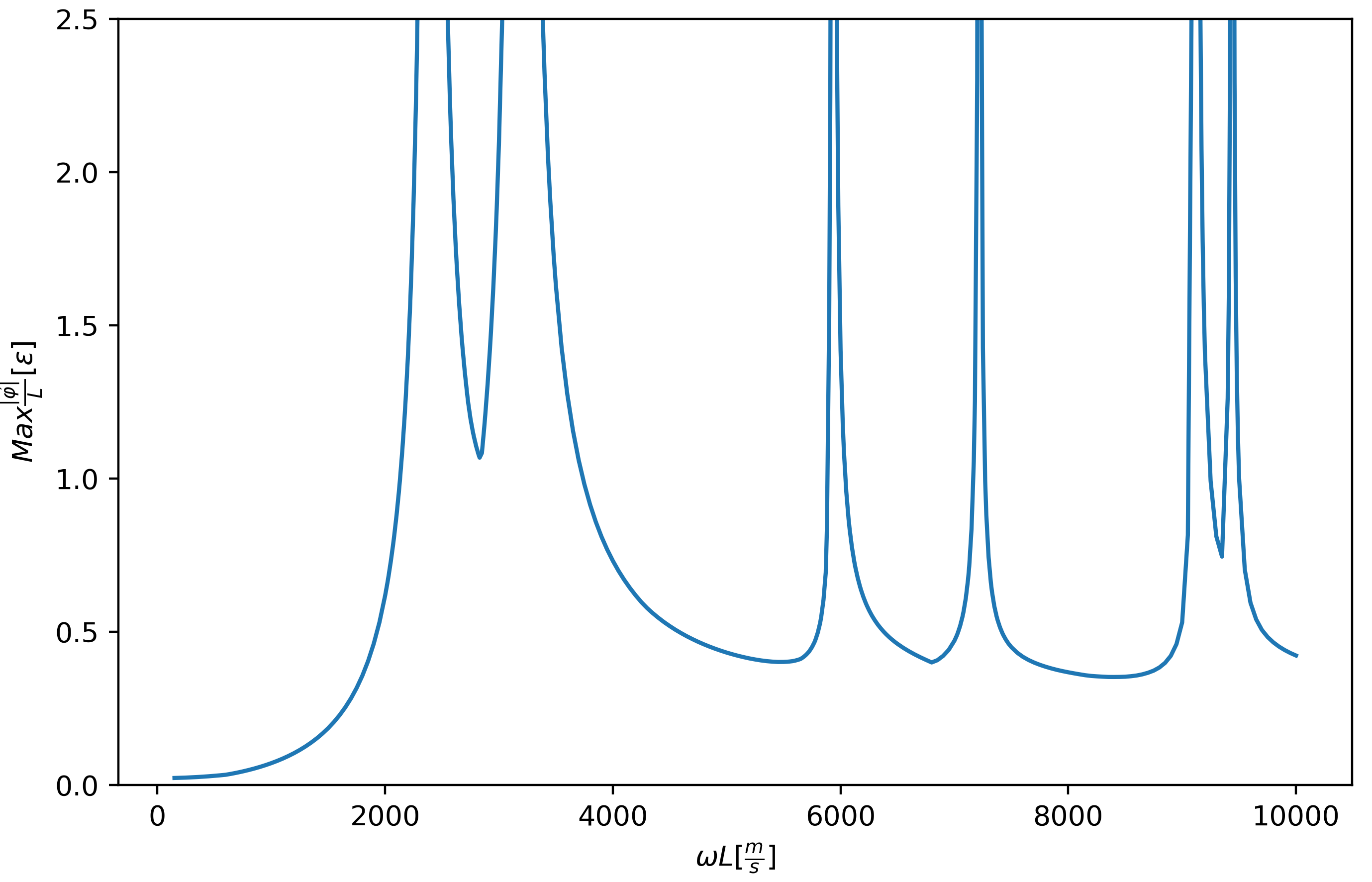}
    \caption{ Maximum of the rescaled physical displacement field $|\varphi|/L$ in units of $\epsilon$ as a function of $\omega L$ for the response to a purely plus-polarized GW.}
    \label{fig:pureAplusPhysicalResonance}
\end{figure}

We discuss only the response to a purely plus-polarized wave shown in Figure \ref{fig:pureAplusPhysicalResonance}, as only this case gives a signal in the interferometer discussed in the next section.

At some frequencies the solution becomes very large, i.e.\ the GW hits a resonance. The frequencies of the first four of these resonances are $\omega_1 \approx \SI{2400}{\hertz}$, $\omega_2 \approx \SI{3200}{\hertz}$, $\omega_3 \approx \SI{5940}{\hertz}$ and $\omega_4 \approx \SI{7220}{\hertz}$.

\begin{figure} 
    \centering
    \subfloat[Vector plot] {{\includegraphics[scale=0.55]{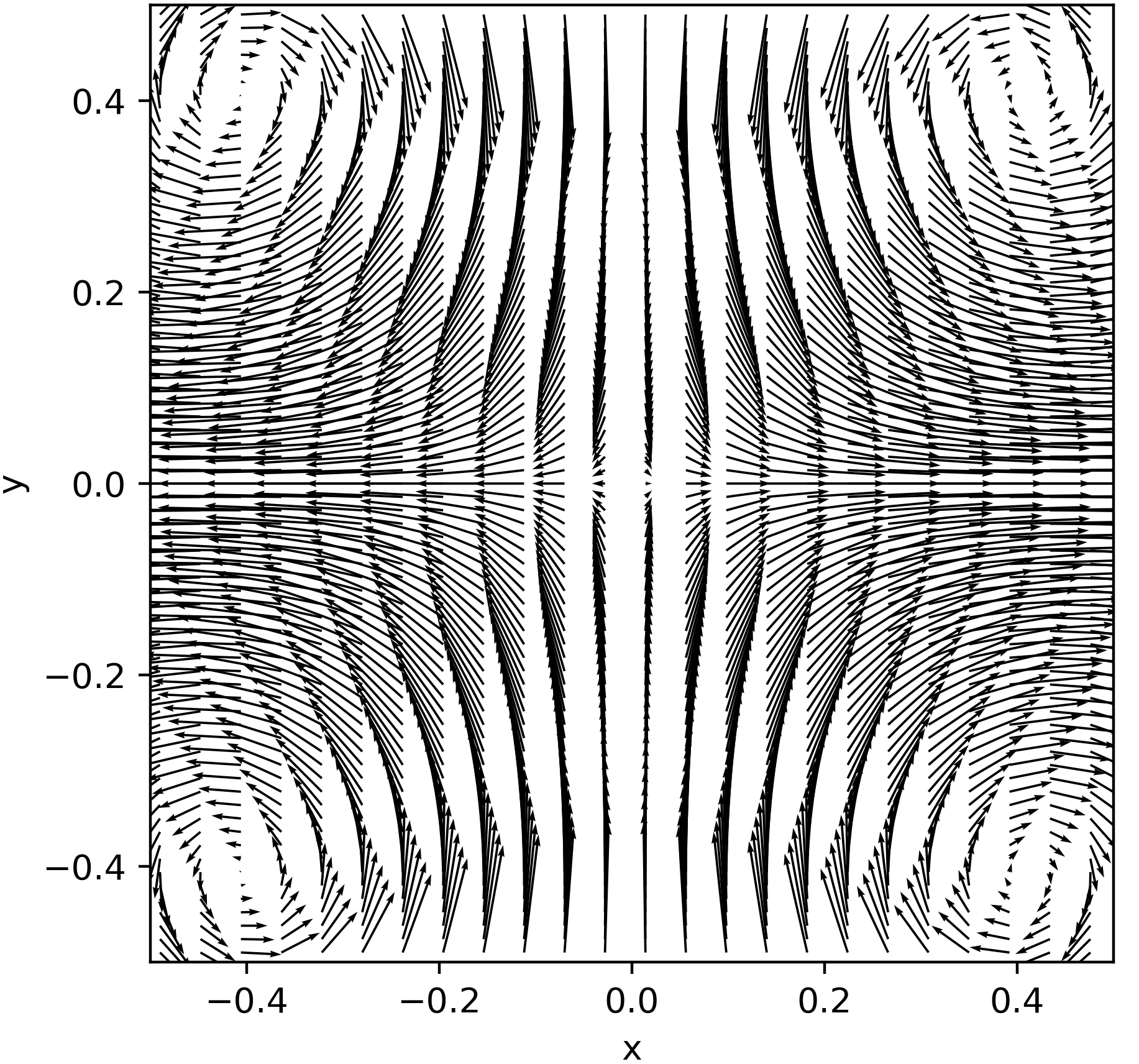}} }
    \qquad
    \subfloat[Grid Plot]{{\includegraphics[scale=0.55]{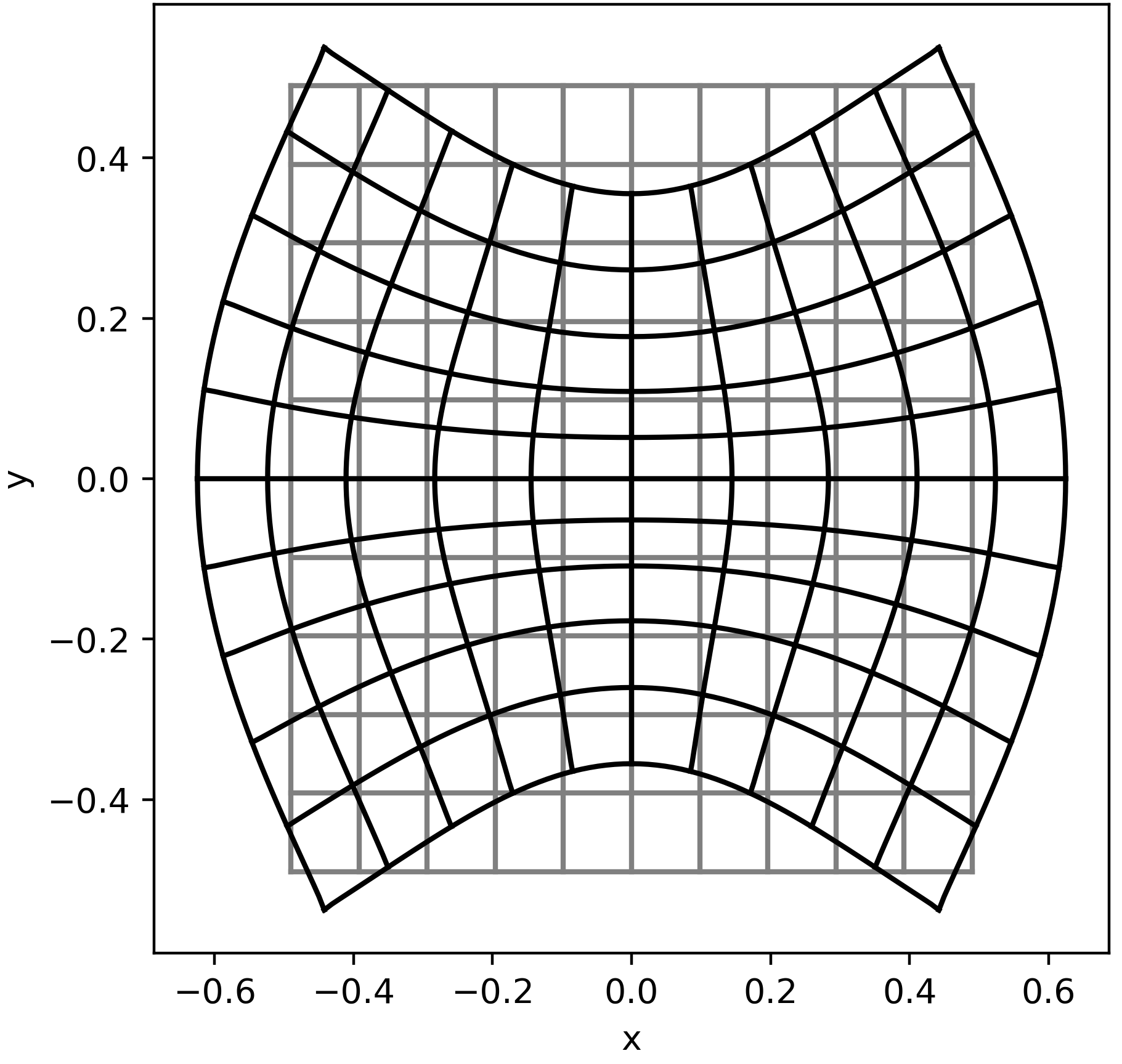} }}
    \caption{ Vector plot (a) and grid plot (b) of the physical displacement field at $\omega= \SI{2700}{\hertz}$ slightly above the first resonance near $\omega_1 \approx \SI{2400}{\hertz}$.}
    \label{fig:pureAplusPhysicalArrowsGrid2700}
\end{figure}

The first resonance looks like the $n=1$ mode of the plate, discussed in \cref{sWaveEigenmodes}. It has a frequency of $\omega_1 = (\SI{540}{\meter\per\second})  \sqrt{2} \pi /\si{\meter} \approx \SI{2399}{\hertz}$.
This agrees very well with the first resonance frequency, which is what is to be expected in a model without damping. A damping term would shift the resonance frequencies.

We would expect to find the $n=2$ mode at the frequency $\omega \approx \SI{4800}{\hertz}$, but there is no corresponding resonance in \cref{fig:pureAplusPhysicalResonance}. 

Following this argument further, the next mode with $n=3$ is at $\omega = \SI{7200}{\hertz}$ and indeed there is a corresponding spike in the resonance curve Figure \ref{fig:pureAplusPhysicalResonance}. This suggests, that only the odd-numbered modes are compatible with the plus-polarized GW. 

We hypothesize that the excitability is related to the mass quadrupole moment, or more precisely to its second time-derivative, as is the case for the emission of gravitational waves via Einstein’s famous quadrupole formula (see e.g.\ \cite{ElementsGR}). One notices that this quantity is non-vanishing only for the odd-numbered modes, which fits nicely with our results.

As each of the resonances in \cref{fig:pureAplusPhysicalResonance} should be related to a normal mode, we expect that there are many more than the ones found in \cref{sec:Eigenmodes}.

\section{Interferometer}

So far, we have only looked at the resonances where the maximum of $\varphi^i$ diverges. These do not necessarily correspond to the frequencies at which the interferometer gives the strongest signal, as it might be possible that the deformations along the path of the laser cancel each other.

\subsection{Interferometer Setup and Assumptions}

\begin{figure}
    \centering
    \includegraphics[scale=0.7]{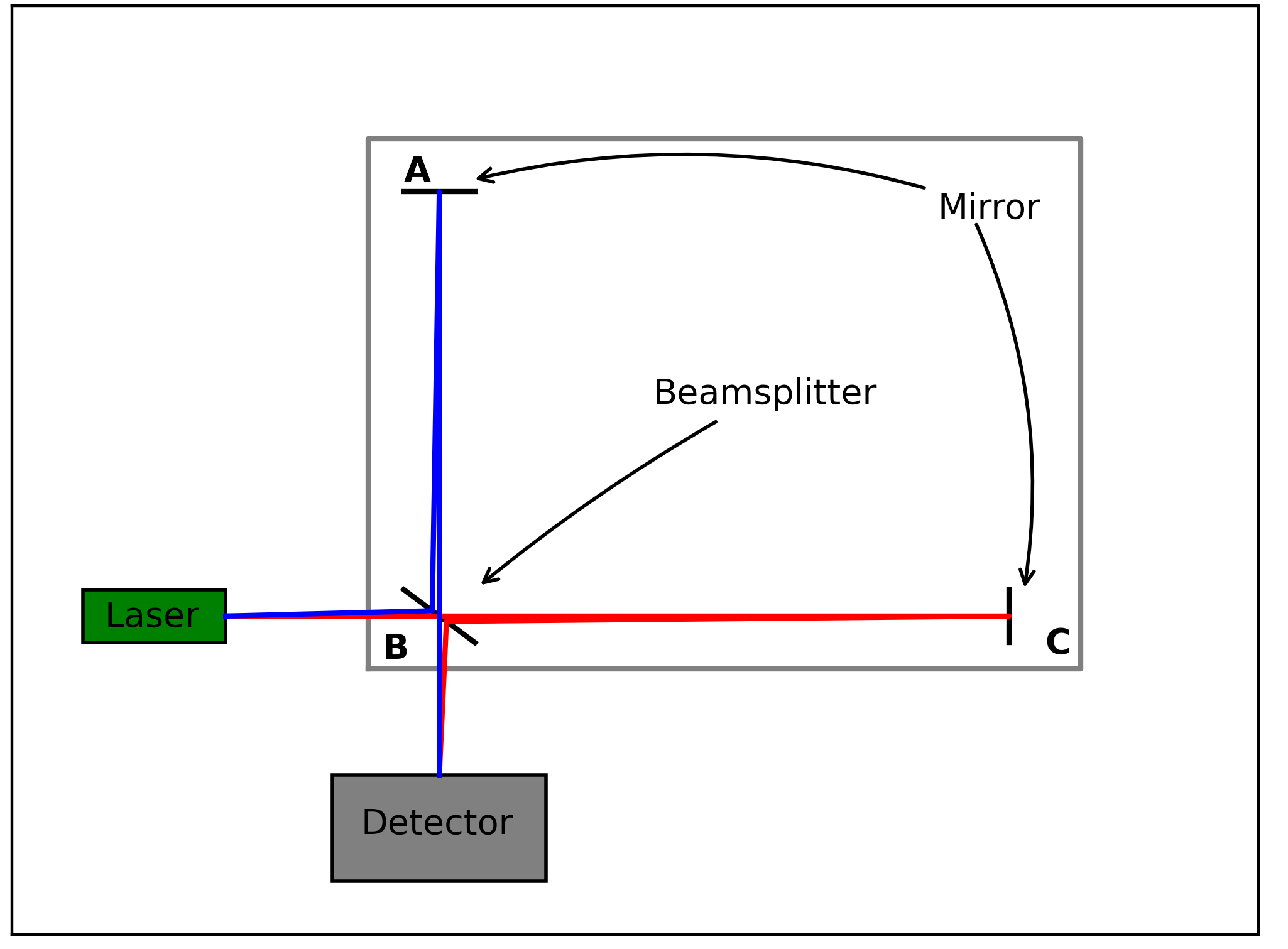}
    \caption{Sketch of an interferometer on a plate consisting of a laser, a beam-splitter, two mirrors and a detector.}
    \label{fig:Interferometer}
\end{figure}
We consider a standard Michelson interferometer, see \cref{fig:Interferometer} and the explanation below. We assume that the instruments are rigid and move as the plate below them does. The phase difference picked up by the laser along the two different paths can be measured as a change in light intensity at the detector. In a more realistic setting, a Fabry--Pérot interferometer would be used, where the laser effectively bounces back and forth multiple times.

For simplicity, it is assumed that the plate does not change while the light crosses the instrument. The light takes about $T_l= 2L/c \approx \SI{e-7}{\second}$ for a $\SI{15}{\metre}$ long plate. The upper limit of frequencies of gravitational waves from known physical phenomena we want to detect is $\SI{10}{\kilo\hertz}$, which corresponds to a period of $T_{GW}= \SI{e-4}{\second}$. For these values, their ratio is $10^{-3}$, so that the assumption of a static plate during the photon flight is justified. This ratio stays constant as long as the product $\omega L$ is constant. Thus, for smaller plates, larger frequencies can be considered, and the other way around. 

The positions of the mirrors and the beam-splitter are $A=(-X_0,Y_0)$, $C=(X_0,-Y_0)$ and  $B=(-X_0,-Y_0)$ (see \cref{fig:Interferometer}). Later, $X_0=Y_0=0.4 L$ is chosen for the calculations. This is to avoid the error at the boundaries due to the finite number of terms in the Fourier series approximation which is largest near the rim of the plate.

\subsection{Calculation of Signal}

In this section, we calculate the phase difference between the laser going along the two different interferometer arms. The calculations are similar to those for the case of free mirrors, found for example in \cite{Rakhmanov} or \cite{melissinos2010response} . We discuss only the case of pure plus-polarization, $A_+=1, A_\times=0$. Looking at the metric in the local Lorentz frame as given in \cite{Rakhmanov}
\begin{equation}
    g_{\mu \nu}=\eta_{\mu \nu}-2
    \begin{pmatrix}
        \Phi & 0 & 0 & \Phi \\
        0 & 0 & 0 & 0 \\
        0 & 0 & 0 & 0 \\
        \Phi & 0 & 0 & \Phi
    \end{pmatrix}\,,
\end{equation}
with $\Phi$ given by
\begin{equation}
    \Phi
        = -\frac{1}{4} \epsilon \ddot{h}_+(t)\left(\xi^{2}-\eta^{2}\right)
        = \frac{\epsilon \omega^2}{4} \cos(\omega t) \left(\xi^{2}-\eta^{2}\right) \,.
\end{equation}
One can see that the $g_{00}$ component is different from the flat-space metric. As it is dependent on the position on the plate, clocks run at different rates depending on where they are. Combining this with the motion of the mirrors, there are two effects that create a phase difference along the two interferometer arms:
\begin{itemize}
    \item A difference in time elapsed because one path is longer than the other.
    \item A difference due to clocks running at different speeds along both paths and hence the light traveling at different coordinate speeds.
\end{itemize}
We calculate the elapsed time along null geodesics (which are still straight lines) taken by the photons. First, we look at the photon moving along the lower interferometer arm from point B to C. For the four-velocity of the photons $K^\mu$ we have
\begin{equation*}
    0
        = g_{\mu \nu} K^{\mu} K^{\nu}
        = (\eta_{tt} - 2 \Phi) (K^t)^2 + \eta_{\xi\xi} (K^\xi)^2
        = -(1 + \tfrac{\epsilon \omega^2}{2} \cos(\omega t) \left(\xi^{2}-\eta^{2}\right) ) (K^t)^2 +  (K^\xi)^2 \,.
\end{equation*}
Using $K^t = \tfrac{dt}{d\lambda}$ and $K^\xi = \frac{d \xi}{d\lambda} $ for a path parameterized by $\lambda$, this can be rearranged to an expression for the coordinate speed of light
\begin{equation}
    \frac{d \xi}{dt}
        = \frac{K^\xi}{K^t}
        = \sqrt{1+ \tfrac{\epsilon \omega^2}{2} \cos(\omega t) \left(\xi^{2}-\eta^{2}\right) }
        \approx 1 + \tfrac{\epsilon \omega^2}{4} \cos(\omega t) \left(\xi^{2}-\eta^{2}\right) \,.
\end{equation}
We again use the assumption that the plate is stationary during one light crossing, i.e.\ the cosine is approximately constant and we take it to be 1 for the maximal possible effect. Integrate along the path from $\xi_B=-0.4 L + \epsilon \varphi^\xi (-0.4 L, -0.4 L)$ to $\xi_C=0.4 L + \epsilon \varphi^\xi (0.4 L, -0.4 L) $ and $\eta = -0.4L$ we find
\begin{align}
    \Delta t_{BC}
        =  0.8 L
        + \epsilon \left[\varphi^\xi (0.4 L, -0.4 L) - \varphi^\xi(-0.4 L, -0.4 L) \right]
        - \epsilon \tfrac{ \omega^2}{4} (0.4 L)^3 (\tfrac{2}{3}  - 2 ) \,.  
\end{align}
The first term is the time it would take the laser to cross the path without any GW. The second term is due to the motion of the mirrors and the last term is the correction due to the different rates at which time passes.

Doing this also for the other interferometer arm the difference in the two elapsed times is found to be given by
\begin{equation}
    \Delta t = \Delta t_{BAB} - \Delta t_{BCB} = \Delta l - \epsilon  \tfrac{4 \omega^2}{3} (0.4 L)^3 \,,
\end{equation}
where $\Delta l$ is the difference in path lengths given by
\begin{equation}\label{SignalExpression}
    \Delta l 
        = 2 \epsilon \cos(\omega t) (
            \varphi^\eta|_A 
            - \varphi^\eta|_B 
            - \varphi^\xi|_C 
            + \varphi^\xi|_B
        ) \,.
\end{equation}
The two effects are additive and can be considered separately. We will later see, that $\Delta l$ is of magnitude $1\epsilon$. The second effect would only be relevant if the plate would be more than 30 kilometers long, in which case our approximations no longer hold. This agrees with the conclusion of \cite{melissinos2010response}. Thus the relevant part of the signal results from the difference in lengths along the two interferometer arms $\Delta l$.

\subsection{Results} 
\label{sec:PhysicalResults}

\begin{figure} 
    \centering
    \includegraphics[scale=0.7]{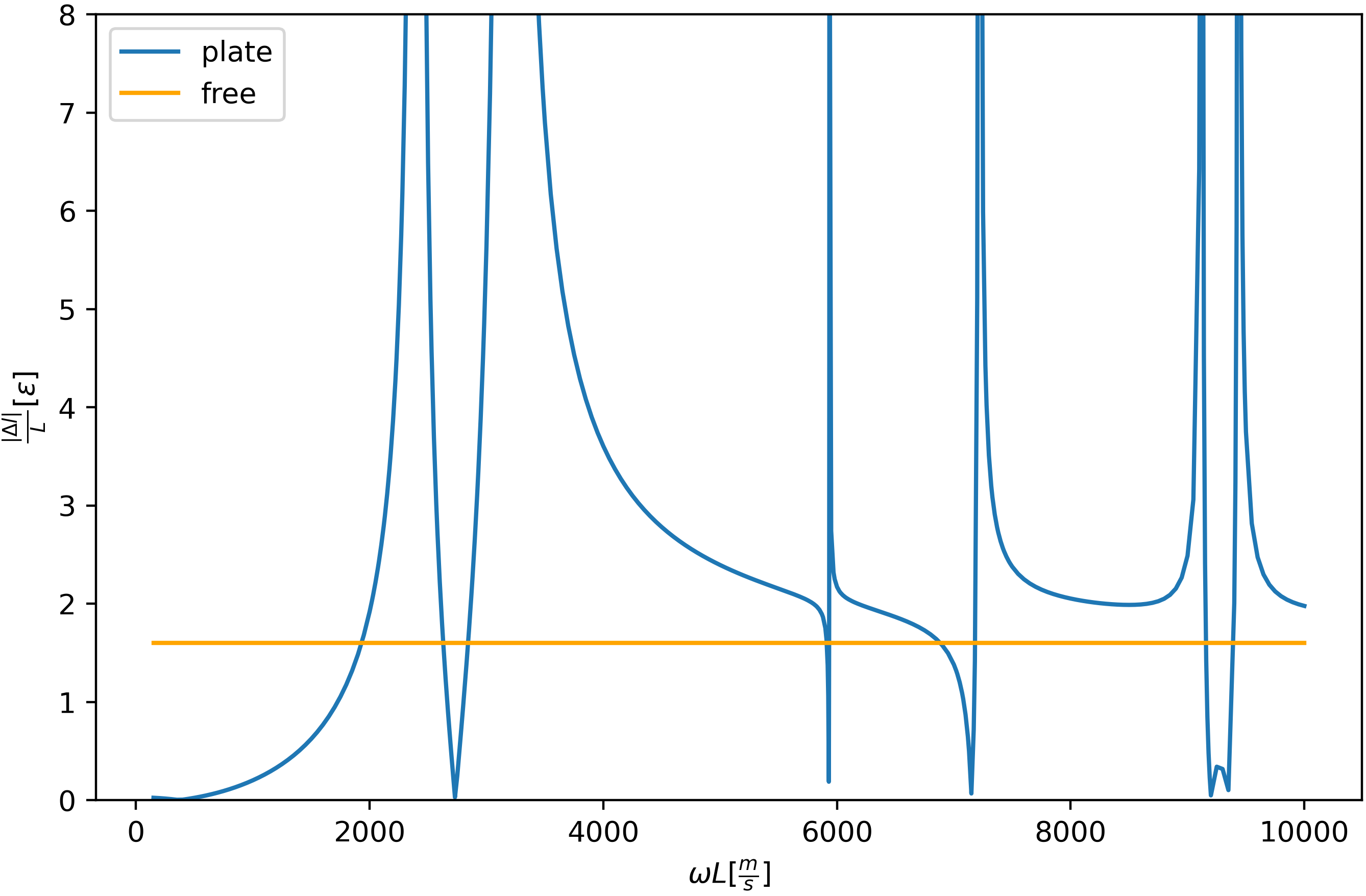}
    \caption{ Difference in relative path lengths $\Delta l$ in units of $\epsilon$ for the interferometer on the plate and the freely suspended mirrors. }
    \label{fig:pureAplus_signal_l1}
\end{figure}

The resulting differences in path lengths for different frequencies are shown in Figure \ref{fig:pureAplus_signal_l1}. Also, for comparison, the resulting signal for an interferometer with freely suspended mirrors (like LIGO,  $\varphi^\xi = \varphi^\eta = 0$) of the same size is plotted. One can see, that there are long ranges (for instance from $\omega L = \SI{3000}{\metre\per\second}$ to almost $ \SI{6000}{\metre\per\second}$) where the signal from the interferometer on the plate is larger. So this is not only the case near the resonances. 

We can choose the size of the plate $L$ so that the frequency range with the largest signal coincides with the range where interesting phenomena are expected. 

\begin{figure}[hb]
    \centering
    \includegraphics[width=0.5\textwidth]{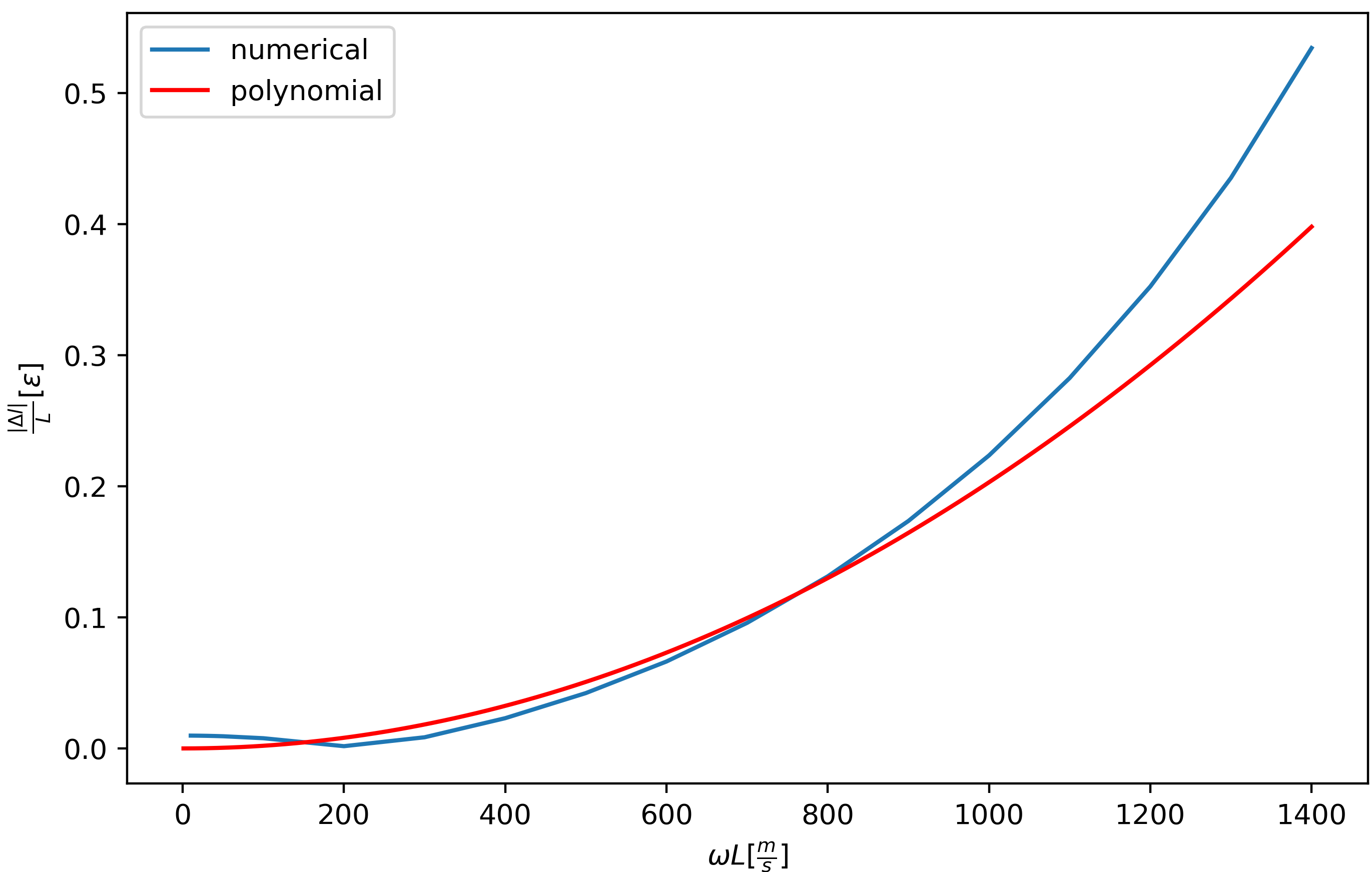}
    \caption{Comparison of polynomial solution in the low-frequency limit and numerical solution from the corrected spectral method.}
    \label{fig:lowFrequencyComparison}
\end{figure}

An approximate polynomial solution for the case of small period ratio $\varepsilon=\tfrac{\omega L}{c_1}$ and pure plus-polarization is given by:
\begin{equation}\label{eq:ApproxAplusSol}
    \varphi^x(x,y) = \tfrac{L}{2} A_+ \left( - x + \tfrac{1}{24}  \varepsilon^2 \tfrac{c_1^2}{c_2^2} \tfrac{c_1^2}{c_3^2} \left[-\tfrac{3}{2} x  \left(  1 - \tfrac{c_2^2}{c_1^2} \right)  + 3 \left( 1-2 \tfrac{c_2^2}{c_1^2} \right)  x y^2 +  x^3 \right] \right) \,.
\end{equation}
Upon conversion of this result to LL coordinates, this can be used to explore the small frequency case (small compared to $\tfrac{c_1}{L}$). The resulting path-length difference is:
\begin{align*}
    \Delta l =  \epsilon L^3 \cos(\omega t) \omega^2 2.03 \times 10^{-7} \,.
\end{align*}
The approximation is valid up to at most $\omega L = 100$ and if we insert this value, we find $\Delta l/L = \epsilon \cos(\omega t) 2.03\times 10^{-3}$. This confirms that the signal indeed almost vanishes for low frequencies. Figure \ref{fig:lowFrequencyComparison} shows a plot of the signal for the polynomial solution and the numerical approach. It can be seen, that the two curves start to diverge at approximately $\omega L = 800$, which is to be expected since the low-frequency limit no longer applies there. Also, for very small frequencies the agreement is not perfect. This is because the numerical approach becomes ill-conditioned. Overall, the two curves become closer, if the Fourier series in the numerical approach are truncated at larger mode number $M$. 

\section{Conclusion}

In this work, we were able to find a numerical solution describing the behavior of a quadratic elastic plate under the influence of a gravitational wave. This was done by developing a spectral approach that deals with the derivatives of Fourier series of non-periodic functions. The validity of this approach was checked in the low-frequency limit against an approximate polynomial solution. 

This solution was then used to calculate the signal a laser interferometer placed on this plate would see. It was discovered, that for broad frequency ranges, the signal is larger than the one an interferometer of the same size, but consisting of freely suspended mirrors, would measure. Of course, the advantage of interferometers with free mirrors, like LIGO, is that they can use kilometer-long tunnels to get a larger signal. Plates are restricted to far smaller sizes. 
Intuitively, one would expect, that the material would oppose the motion of the mirrors and thus minimize the signal. As demonstrated above, this is only true for gravitational wave frequencies which are small compared to the first resonance frequency of the plate.

The signal is especially large near certain resonance frequencies. Looking at the motion pattern of the plate near the resonances, we discovered that some of them correspond to the normal mode solutions from section \ref{sWaveEigenmodes}. 
The amplitude of the signal is of the order of $\epsilon L \approx 10^{-20} L$. This can be improved by using a larger plate or longer effective path lengths with the help of a Fabry--Pérot interferometer to bounce the laser back and forth multiple times. Though there are limits to these improvements, as the effects will start to cancel once the plate undergoes more than half of its oscillation during the light-crossing. The resulting path-length difference has to be compared to the wavelength of the laser which is of order $\SI{e-7}{\meter}$. As this is far larger, the phase difference and hence the interference effect will be very small.

To get an idea of the practicability of these measurements, one would need to compare the resulting deformations to the ones caused by thermal noise and seismic disturbances of the plate. Moreover, our model does not include any damping behavior, which would also reduce the amplitude of the signal and make experimental observations even more difficult.

\section*{Acknowledgements}
We are grateful to Piotr Chruściel and Robert Beig for many helpful discussions.
T.M. is a recipient of a DOC Fellowship of the Austrian Academy of Sciences at the University of Vienna, Faculty of Physics, and is supported by the Vienna Doctoral School in Physics (VDSP), the research network TURIS, and in part by the Austrian Science Fund (FWF), Project No.~P34274, as well as by the European Union (ERC, GRAVITES, 101071779).

\appendix{}

\section{Bulk Solutions}
\label{sec:FourierAnsatz}

With a plane wave ansatz
\begin{equation}
    \varphi^j(x,y) = a^j e^{i k_l x^l} \,,
\end{equation}
the equations of motion reduce to the matrix system
\begin{equation}
    B \vec{a} =
    \begin{pmatrix}
            (k^2 c_2^2-\omega^2) + c_3^2 k_x^2 & c_3^2 k_x k_y \\
            c_3^2 k_x k_y & (k^2 c_2^2-\omega^2) + c_3^2 k_y^2
    \end{pmatrix}
    \begin{pmatrix}
            a^x \\ a^y
    \end{pmatrix} = 0 \,.
\end{equation}
For non-zero $a$, the determinant of $B$ must vanish.
There are two values of $k_j$, and the corresponding 0-Eigenvectors of $B$, for which this is the case:
\begin{itemize}
    \item p-waves: $\omega=c_1 \kappa$
        where the wave vector is denoted $\kappa_j$ to distinguish it from the second case.
        The matrix then simplifies to
        \begin{equation*}
            \frac{\mu+\lambda}{\rho_0}
            \begin{pmatrix}
                -\kappa_y^2 & \kappa_x \kappa_y \\
                \kappa_x \kappa_y & -\kappa_x^2
            \end{pmatrix}
            \begin{pmatrix}
                    a^x \\ a^y
            \end{pmatrix} = 0 \,,
        \end{equation*}
        and is solved by vectors $a_j$ which are collinear with $\kappa_j$, i.e. longitudinal oscillations.
    
    \item s-waves: $\omega=c_2 k$
    In this case we call the amplitude $b^i$ instead of $a^i$ to easier distinguish between the two cases.
    Here the matrix becomes
        \begin{equation*}
            \frac{\mu+\lambda}{\rho_0}
            \begin{pmatrix}
                k_x^2 & k_x k_y \\
                k_x k_y & k_y^2
            \end{pmatrix}
            \begin{pmatrix}
                    b^x \\ b^y
            \end{pmatrix} = 0 \,
        \end{equation*}
        and is solved when $k_j b^j = 0$, i.e. transversal oscillations. 
\end{itemize}

\section{Normal Modes in Flat Space}
\label{sec:Eigenmodes}

Without a GW ($h_{ij}=0$), the differential equations we want to solve stay the same, but in the boundary conditions \eqref{eq:BoundaryConditionsPhi} the right-hand side vanishes. 

\subsection{S-Wave Modes}
\label{sWaveEigenmodes}

\begin{figure}[h]
    \centering
    \subfloat{\includegraphics[scale=0.5]{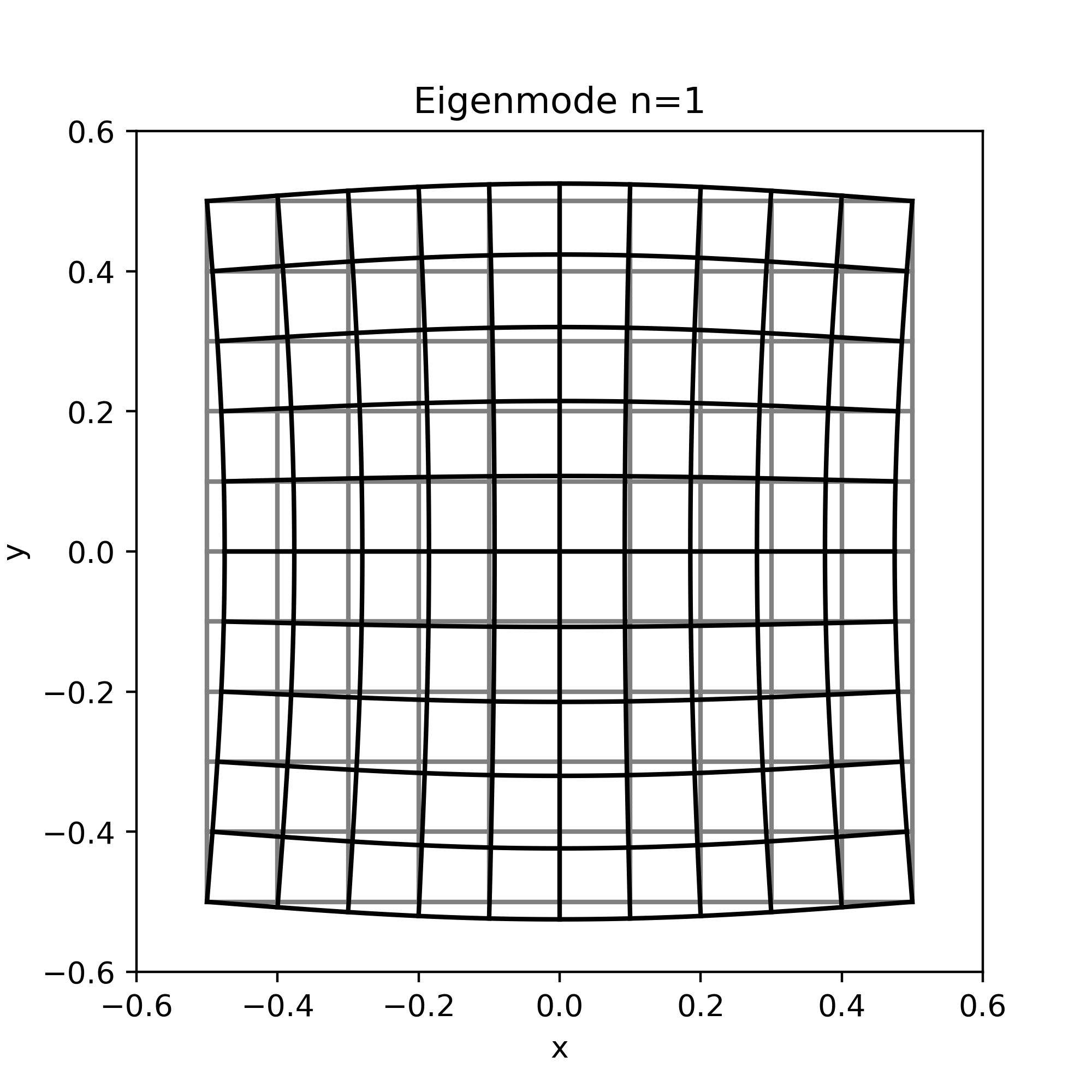}}
    \qquad
    \subfloat{\includegraphics[scale=0.5]{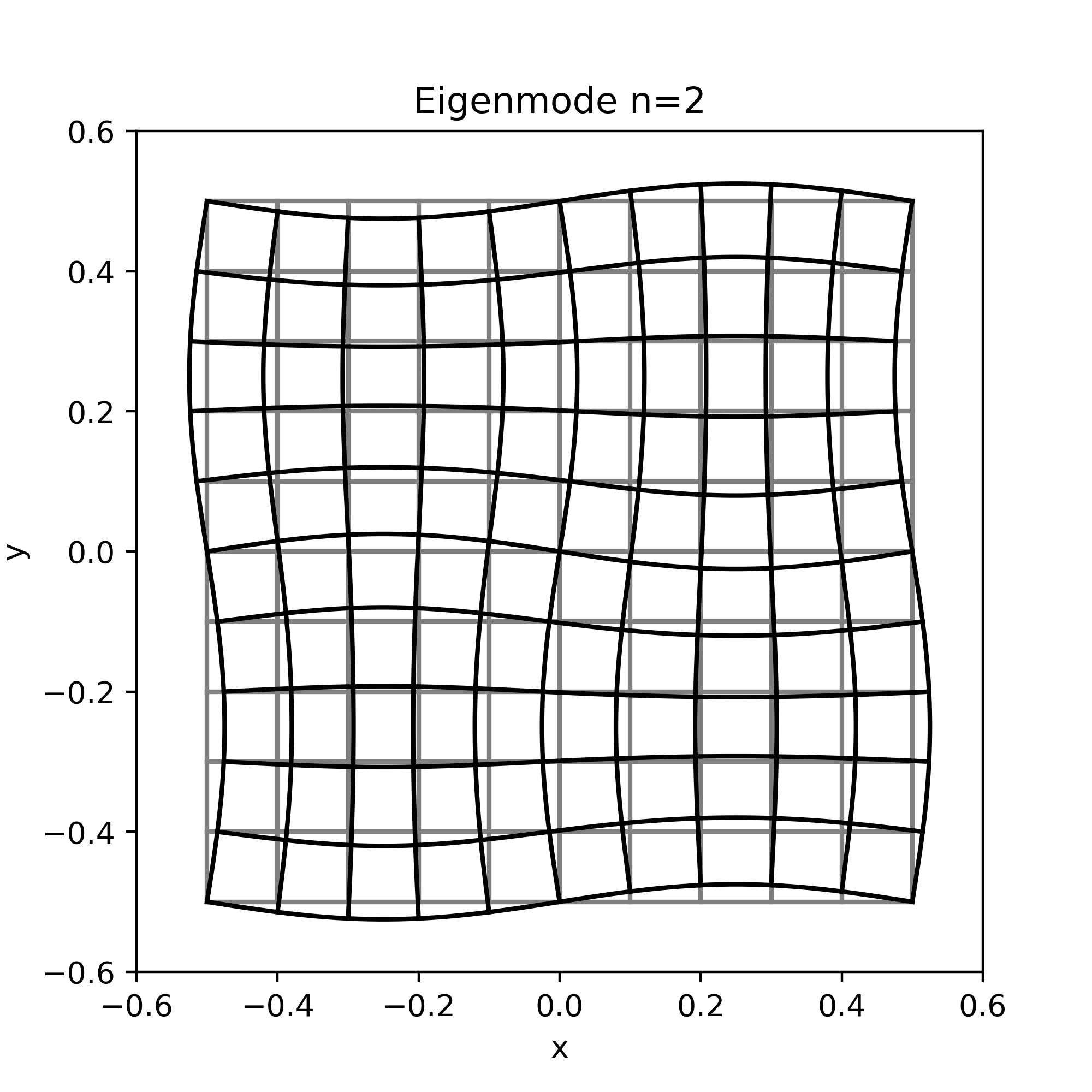}}
    \caption{Grid plots of the first two normal modes. The grey lines represent the undisturbed plate, the black lines the deformed one. The deformation is not to scale.}
    \label{fig:my_label}
\end{figure}

Here the solutions to the PDE have the form $\varphi^j=b^j \ e^{ik_l x_l}$ with $k_j b^j  = 0$. 
It might be the case that a combination of more such solutions is needed to satisfy the boundary conditions. For the separation of the time dependence to still work, they all need to have the same magnitude of $\vec{k}$ (and therefore also $\omega$), while the direction is still free. Which $k_j$-vectors should we combine? Looking at one boundary, e.g.\ $x=\tfrac{L}{2}$, we have a linear combination of $\cos(k_y y)$  and $\sin(k_y y)$ which has to vanish. The only other $k_j$ which can help to cancel these terms are the ones with the same $k_y$. The same follows for $k_x$ from the condition on the boundary $y=\tfrac{L}{2}$. So only two $k_j$ vectors can be combined helpfully:
\begin{equation}
    k_j^{(1)} = (k_x, k_y) \text{ and } k_j^{(2)} = (k_x, -k_y) \,.
\end{equation}
Putting this all together $\varphi^i$ can be written in terms of cosines and sines as
\begin{align} \begin{split} \label{eq:GeneralSWave}
     \varphi_x &= k_y \left( A c_x c_y + B s_x s_y + C s_x c_y + D c_x s_y \right) \,, \\
     \varphi_y &= k_x \left( B c_x c_y + A s_x s_y - D s_x c_y - C c_x s_y \right) \,,
\end{split}\end{align}
where the notation $\cos(k_x x) = c_x$ and similar has been used. Since $\vec{\varphi}$ is divergence-free, the boundary conditions reduce to
\begin{equation}
\begin{aligned}[t]
    \sigma_{xx}\rvert_{\partial P_x} & = \epsilon 2 c_2^2 \partial_x  u^x  & = 0 \,, \\
    \sigma_{yy}\rvert_{\partial P_y} & = \epsilon 2 c_2^2  \partial_y  u^y  & = 0 \,, \\
    \sigma_{xy}\rvert_{\partial P} & = \epsilon c_2^2[ \partial_x  u^y +\partial_y  u^x ] & = 0 \,.
\end{aligned}
\end{equation}
Evaluating the first condition with the help of Mathematica for $\partial P_x$ at $(\pm \frac{L}{2}, \pm y)$ and adding the four terms with different sign combinations ($++++$, $++-+$, $+-+-$, $+---$) gives four simpler, necessary but not sufficient conditions:
\begin{equation}
    C k_x k_y   \cos \left(\frac{k_x L}{2}\right) \cos \left(k_y y\right) = 0 \,,
\end{equation}
\begin{equation}
    B k_x k_y   \cos \left(\frac{k_x L}{2}\right) \sin \left(k_y y\right) = 0 \,,
\end{equation}
\begin{equation}
    A k_x k_y   \sin \left(\frac{k_x L}{2}\right) \cos \left(k_y y\right) = 0 \,,
\end{equation}
\begin{equation}
    D k_x k_y   \sin \left(\frac{k_x L}{2}\right) \sin \left(k_y y\right) = 0 \,.
\end{equation}
Doing something similar with the other conditions/boundaries and trying to solve them all, results in two possible mode solutions:
\begin{equation}
\begin{aligned}
    A=B=C=0 &,\ \ \ k_x=k_y=\frac{2 \pi}{L} n \,, \\
    A=B=D=0 &,\ \ \ k_x=k_y=\frac{\pi}{L} (2n+1) \,.
\end{aligned}
\end{equation}
Therefore the whole solution $\vec{u}$ can be written as follows.
\begin{mdframed}[backgroundcolor=gray!40]
\textbf{“Quadratic” S-Wave Modes:}
\begin{itemize}
    \item For $n$ even:
    \begin{align}
        u^x & =  + \cos(\omega t) \ \cos\left( \frac{\pi}{L} n x\right) \ \sin\left( \frac{\pi}{L} n y\right) \,, \\
        u^y & = -\cos(\omega t) \  \sin\left( \frac{\pi}{L} n x\right) \ \cos\left( \frac{\pi}{L} n y \right) \,.
    \end{align}
    \item For $n$ odd:
    \begin{align}
        u^x & = - \cos(\omega t) \ \sin\left( \frac{\pi}{L} n x\right) \ \cos\left( \frac{\pi}{L} n y \right) \,. \\
        u^y & = +\cos(\omega t) \  \cos\left( \frac{\pi}{L} n x \right) \ \sin\left( \frac{\pi}{L} n y \right)\,.
    \end{align}
\end{itemize}
\vspace{-1\baselineskip}
\end{mdframed}
\Cref{fig:my_label} shows the modes for $n=1$ and $n=2$. For these modes, the corners always stay fixed.

\subsection{Mixed Modes}

So far, we saw that there are s-wave modes and want to explore if there are p-wave or mixed modes. This is answered by the following theorem.
\begin{theorem}
    Consider solutions $\varphi^j$ made up of the building blocks  $b^j e^{ik_l x^l}$, $k_j b^j= 0$ (s-waves) and $a^j e^{i \kappa_l x^l}$, with $a^l \parallel \kappa^l$ (p-waves). The magnitude of the wave vectors is given by the dispersion relations $\omega=c_1 \kappa$ and $\omega =c_2 k$ while the direction is still free to choose.
    Then, there exists no finite sum of such terms which satisfies the boundary conditions \eqref{eq:BoundaryConditionsPhi} with vanishing right-hand side, except for pure s-wave modes \eqref{sWaveEigenmodes}.
\end{theorem}
As mentioned above, we need only consider a fixed value of $\omega$.
\begin{proof}
    First, note that since $c_1 > c_2$ the relation $\kappa < k$ for the magnitudes of the wave vectors holds. Since we do not want a pure s-wave solution, $\varphi^j$ contains at least one p-wave term $a^j \ e^{i \kappa_l x^l}$. The most general p-wave solution can be written as 
    \begin{align}\begin{split}
        \varphi_x &= \kappa_x \left( A c_x c_y + B s_x s_y + C s_x c_y + D c_x s_y \right) \,, \\
         \varphi_y &= \kappa_y \left( -B c_x c_y - A s_x s_y + D s_x c_y + C c_x s_y \right) \,,
    \end{split} \end{align}
    for some constants $A,B,C,D$.
    This leads to a $\sigma_{xx}$ component, evaluated at the $x=\frac{1}{2}$ boundary which has the following form
    \begin{equation}
        \sigma_{xx}\left(\frac{L}{2},y\right)= E \cos(\kappa_y y) + F \sin(\kappa_y y) \,,
    \end{equation}
    with some constants $E$ and $F$. It does not satisfy the boundary condition on its own, so we need to add an s-wave term with the same y-component $k_y=\kappa_y$. But the $k_x$ component of the s-wave is then given by
    \begin{equation}
        k_x= \pm \sqrt{\left(\frac{\omega}{c_2}\right)^2 - k_y^2} \,.
    \end{equation}
    As $|k_j| > |\kappa_j|$ there is no other possible p-wave with the same $\kappa_x$ component, so this term has to satisfy the boundary conditions on $\partial P_y$ on its own, in particular:
    \begin{equation}
        \sigma_{yy}\left(x,\frac{L}{2}\right) \pm \sigma_{yy}\left(x,-\frac{L}{2}\right) = 0 \,.
    \end{equation}
    The general form of such a s-wave then looks like \eqref{eq:GeneralSWave} and if we insert the resulting CS-tensor in the above relation two necessary conditions follow:
    \begin{align*}
    k_x k_y   \cos \left(\frac{k_y L}{2}\right) \left(A \sin \left(k_x x\right)-C \cos \left(k_x x\right)\right) &= 0 \,, \\
    k_x k_y   \sin \left(\frac{k_y L}{2}\right) \left(D \sin \left(k_x x\right)-B \cos \left(k_x x\right)\right) &= 0 \,.
    \end{align*}
    This has to be true for all $x$. Since $k_j$ is already fixed, only the constants $A, B, C$ and $D$ can be used to satisfy these conditions. But in general, the only possibility to do this is $A=B=C=D=0$. Therefore there is no s-wave and we are back at a pure p-wave. But we already saw in the previous section that this does not satisfy the boundary conditions.
\end{proof}
This concludes our analytic investigation of mode solutions. In terms of finite sums, there are only the quadratic s-wave modes \eqref{sWaveEigenmodes}. To find more general solutions the whole Fourier series would need to be considered.

\bibliographystyle{abbrv}
\bibliography{bibliography}

\end{document}